\documentclass{ws-procs975x65}

\begin{document}

\title{DO SGRs/AXPs AND RADIO AXPs HAVE THE SAME NATURE?}

\author{GOULART COELHO, JAZIEL$^*$; MALHEIRO, MANUEL}

\address{Departamento de F\'{i}sica, Instituto Tecnol\'{o}gico de Aeron\'{a}utica,\\
S\~{a}o Jos\'{e} dos Campos, SP 12.228-900, Brazil\\
$^*$E-mail: jaziel@ita.br\\
www.ita.br}

%

\begin{abstract}

SGRs/AXPs are assumed to be a class of neutron stars (NS) powered by magnetic energy and not by rotation, as normal radio pulsars. 
However, the recent discovery of radio-pulsed emission in four of this class of sources, where the spin-down rotational energy lost
$\dot{E}_{\rm rot}$ is larger than the X-ray luminosity $L_X$ during the quiescent state - as in normal pulsars - opens the 
question of the nature of these radio AXPs in comparison to the others of this class. In this work, we show that the radio AXPs
obey a linear log-log relation between $L_X$ and $\dot{E}_{\rm rot}$, very similar to the one of normal X-ray pulsars,
a correlation not seen for the others SGRs/AXPs. This result suggests a different nature between the radio AXPs comparing
to the others SGRs/AXPs.
\end{abstract}


\bodymatter

\section{Introduction}\label{aba:sec1}

Over the last decade, observational evidence has mounted that SGRs/AXPs belong to a particular class of pulsars\cite{Mereghetti}.
They are understood in the framework of strongly magnetized neutron star\cite{Thompson,Duncan}, but there are alternative scenarios, in particular
the white dwarf (WD) pulsar model\cite{MMalheiro,Coelho,Boshkayev}. Recently, 
four over a total of about 20 SGRs/AXPs presented radio-pulsed emission. These radio AXPs showed several properties that make them different
from the others SGRs/AXPs: loud radio (transient radio emission different of the radio pulsar emission), low quiescent
X-ray luminosity $L_X$ (decreasing with time) that can be explained from the spin-down rotational
energy lost $\dot{E}_{\rm rot}$ of a neutron star, as normal rotation-powered pulsar. As pointed out in Malheiro et al.\cite{MMalheiro}, 
the X-ray efficiency $\eta_X=L_X/\dot{E}_{\rm rot}$ for these radio AXPs, seems to be to small comparing to the others SGRs/AXPs
when interpreted as magnetized white dwarfs. However, as neutron star pulsars these radio AXPs have
$\eta_X\sim(0.2-0.1)$, a little bit larger than the values of normal pulsars (where $\eta_X\sim(0.3-0.4)$). This can be understood due to
their large magnetic dipole momentum $m\sim10^{32}$ emu,
giving support for their neutron star interpretation.
In this contribution, we suggest that the radio SGRs/AXPs are rotation-powered neutron star pulsars $L_X\simeq k\dot{E}_{\rm rot}^n$ ($L_X<\dot{E}_{\rm rot}^{NS}$),
in contrast to the others that are rotation-powered magnetized white dwarfs.
In our understanding the large steady X-ray luminosity  
seen for almost all the no-radio SGRs/AXPs,
can be explained as coming from a large spin-down 
energy lost of a massive white dwarf with a much large magnetic dipole moment of $10^{34}\leq m\leq10^{36}$
emu consistent with the range observed for isolated and very magnetic WDs (see Coelho \& Malheiro 2012\cite{Coelho2}
for discussions), indicating a different nature between these sources and the radio AXPs.
These radio AXPs have large magnetic
field and seem to be very similar to the high-B pulsars recently founded\cite{kaspi}, as already
pointed out in Ref.\cite{NandaRea} (see Fig.~1 - right panel of this contribution). However, we 
should emphasize that even if the radio AXPs are strong magnetized neutron stars,
they are not magnetars in the sense that their steady luminosity is not originated by the magnetic
energy, but from the
rotational energy as normal pulsars.

\section{Results and Discussions}
Recently, Rea et al. (2012)\cite{NandaRea} try to understand magnetar radio emissions from an phenomenological point of
view. They proposed that magnetars are radio-loud if and only if their quiescent X-ray luminosities are smaller than their rotational
energy loss rate ($L_X< \dot{E}_{\rm rot}$). 
In Fig.~1 - left panel, we see a large steady X-ray luminosity (and almost
constant as a function of $\dot{E}_{\rm rot}$) for almost all the no-radio SGRs/AXPs (black circle points - high $\dot{E}_{\rm rot}$ and high $L_X$), indicating
a different nature among these
sources and the radio AXPs (green square points - high $\dot{E}_{\rm rot}$ and low $L_X$).
These radio AXPs are ordinary pulsars, in line with their steady X-ray luminosity that can be well explained
within the neutron star model (see Fig.~1 - left panel), and with magnetic fields close to the ones of high-B pulsars (see 
Fig.~1 - right panel). We show that the radio AXPs obey a linear log-log relation between $L_X$ and $\dot{E}_{\rm rot}$,
$\log L_X= \log(5.9\times10^4)+0.814\log{\dot{E}_{\rm rot}}$, very similar to the one
satisfied by X-ray and gamma-ray neutron star pulsars\cite{Becker,Vink,Kargaltsev,Abdo,Arons}, suggesting their neutron star nature.
In contrast, for almost all the others SGRs/AXPs, $\log L_X$ does not vary too much as
function of $\log \dot{E}_{\rm rot}$, a phenomenology not shared by X-ray neutron star pulsars, supporting
a different nature for these sources.
\begin{figure}[!htp]
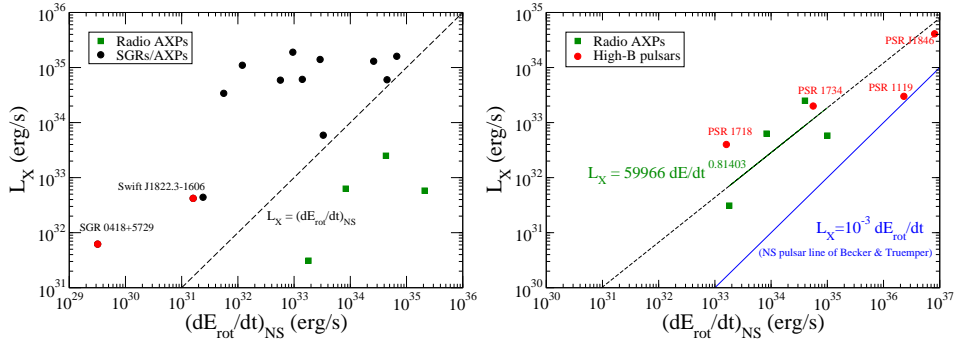

\begin{center}
\psfig{file=LxEdot_NS.eps,width=2.45in}
\psfig{file=LxEdottrans.eps,width=2.45in}
\end{center}
\caption{(Color online) {\it{Left panel}}: X-ray luminosity $L_X$ versus the loss of rotational energy $\dot{E}_{\rm rot}$, describing SGRs and AXPs as
neutron stars. The red points correspond to recent discoveries of SGR 0418+5729 and Swift J1822.3-1606 with low magnetic field. {\it{Right panel}}:
X-ray luminosity $L_X$ versus the loss of rotational energy $\dot{E}_{\rm rot}$ for the four
AXPs that show radio-pulsed emission, together with some high-B pulsars. A linear $\log-\log$ relation between $L_X$ and $\dot{E}_{\rm rot}$ is found
for the radio AXPs, $L_X\propto \dot{E}^{0.81403}$ (green dashed-line), very similar to the X-ray NS pulsar line of Becker \& Tr$\rm \ddot{u}$mper\cite{Becker}.
}
\label{fig1}
\end{figure}

\section{Acknowledges}
The authors acknowledges the financial support of the Brazilian agency CAPES, CNPq
and FAPESP (S\~{a}o Paulo state agency, thematic project $\#$ 2007/03633-3).

\end{document}